\documentclass[aps,prd,amsmath,amssymb,showkeys]{revtex4}
\usepackage{amsfonts}
\usepackage{graphicx}% Include figure files
\usepackage{bm}% bold math

\def\journal#1#2#3#4{{#1} {\bf #2}, #3 (#4)}
\newcommand{\be}{\begin{equation}}
\newcommand{\ee}{\end{equation}}
\newcommand{\bea}{\begin{eqnarray}}
\newcommand{\eea}{\end{eqnarray}}
\newcommand{\hf}{\frac12}
\newcommand{\nn}{\nonumber\\}
\def\eq#1{(\ref{#1})}
\def\la{\langle}
\def\ra{\rangle}
\def\Tr{{\mathrm{Tr}}}

\def\mr#1{{\mathrm{#1}}}
\def\v#1{{\bm{#1}}}
\def\ord#1{{\cal O}(#1)}
\def\br{\hskip-6pt/}

\def\Im{\mr{Im}\hskip1pt}
\def\Re{\mr{Re}\hskip1pt}
\def\etab{\bar\eta}

\def\heta{\hat\eta}
\def\hetab{\hat{\bar\eta}}
\def\hj{\hat j}
\def\hpsi{\hat\psi}
\def\hpsib{\hat{\bar\psi}}
\def\hA{\hat A}
\def\hD{\hat D}
\def\hG{\hat G}

\def\psib{\bar\psi}
\def\ub{\bar u}

\begin{document}
\title{Proton scattering on an electron gas}
\author{Fares Mansouri}
\email{mansouri.fares1@gmail.com}
\affiliation{University of Strasbourg, High Energy Physics Theory Group, CNRS-IPHC, 23 rue du Loess, BP28 67037 Strasbourg Cedex 2, France}
\author{Janos Polonyi}
\homepage{http://www-physique-ingenierie.u-strasbg.fr/~polonyi/}
\affiliation{University of Strasbourg, High Energy Physics Theory Group, CNRS-IPHC, 23 rue du Loess, BP28 67037 Strasbourg Cedex 2, France}
\author{Karima Zazoua}
\email{karima67@yahoo.fr}
\affiliation{University of Strasbourg, High Energy Physics Theory Group, CNRS-IPHC, 23 rue du Loess, BP28 67037 Strasbourg Cedex 2, France}
\affiliation{LEPM USTO-MB, BP 1505 EL M`naouer Oran, Algeria}
\author{Nouredin Zekri}
\affiliation{LEPM USTO-MB, BP 1505 EL M`naouer Oran, Algeria}
\begin{abstract}
It is shown in the case of proton scattering on an electron gas target that the Closed Time  Path formalism can handle final state interactions of the target in equilibrium in a simple and natural manner. The leading order cross section is proportional to the photon density of states. The scattering needs a partial resummation of the perturbation series when the electron gas forms long living quasi-particles with high density of state during the collision. A strong cancellation between real and virtual electron-hole pairs is found in this case.
\end{abstract}
\date{\today}
\pacs{}
\keywords{scattering, final state interaction, environment, infrared divergences}
\maketitle

\section{Introduction}
The adiabatic picture of scattering theory, the assumption of the noninteracting nature of  the colliding particles well before and after the collision, is justified by assuming short range interactions. But there is usually a large number of particles in the target which remain interactive even after the colliding particles have left the collision zone and become well separated. It is natural to consider the colliding particles, tracked by the detectors as system and the target particles as environment. The system-environment entanglement, generated during the collision remains present and may influence the observed transitions rates, observed by well separated detectors. One can naturally take into account this effect by summing the transition probability over all excited environment states in calculating the reduced system density matrix. The main point of the present work is that this sum which is over a large number of state and seems difficult to handle can be carried out in a natural manner within the Closed Time Path (CTP) formalism \cite{schwingere,schwingerk,keldysh,feynman,umezawa,niemisa,calzetta}.

Let us consider for the sake of definiteness the scattering of a proton beam on an electron gas target, described by a homogeneous gas of density $n_g$ and size $\ell_t\to\infty$. The proton beam can be imagined in its rest frame as a gas bounded in momentum space by an anisotropic Fermi surface, with characteristic size given by the average and the dispersion of the momentum of the beam particles. We simplify matters in what follows by assuming a sufficiently dilute and clean beam, rendering its collective excitations negligible. Furthermore, the electron density is supposed to change smoothly at the wall of the container to neglect the change of the beam particle momentum upon entering and leaving the target.

One can separate different time scales in this problem. The initial and final time, $t_i$ and $t_f$ correspond to the time when the asymptotic, free in and out description of scattered particle states applies. A single proton wave packet enters and leaves the target at time $t_{in}$ and $t_{out}$, respectively and we assume the inequalities $i_i\ll t_{in}\ll t_{out}\ll t_f$. Furthermore, the shortest characteristic time scale of the target, $t_{micr}$, the average time between two consecutive collisions is supposed to satisfy the inequality $t_{micr}v\ll\ell_t$ where $v$ denotes the beam velocity, assuring that the beam interacts with the possible collective modes of the target.

It has already been noted in \cite{scattenv} that the summation over the environment final states can easier be carried out by the CTP method. The scattering of a proton on an electron gas was considered in that work for collision processes where the initial electron was at rest within the gas. The generalization of that result is presented here by relaxing the restriction on the initial state of the electron. i.e. by allowing that the proton collides with any particle of the target. We simplify the problem further by considering inclusive scattering for unpolarized particles where the initial and final momentum of the proton are recorded only. It is shown in the present work that such a more realistic treatment of the target particles leads to an even simpler expression of the cross section.

The advantage of the CTP formalism is that one deals with observables, in particular transition probability rather than transition amplitudes. In the usual scheme, based on the transition amplitude one works out two perturbation series, one for the amplitude and another for its complex conjugate and multiplies them together. This product may display a rather involved structure due to cancellations between real and virtual contributions. The CTP formalism offers a graphical representation of the perturbation series of the probability directly where the real and virtual terms appear on equal footing, rendering their calculation and comparison easier.

It is found that the scattering process is strongly coupled when the photon exchanged between the scattering charges is dominated by long living electron-hole quasi-particle pairs and a strong cancellation is found among real and virtual contributions in this case. This is similar to the case of IR divergences where a cancellation is observed between real and virtual photons when their propagators diverge \cite{bloch,yennie}.

The organization of the paper is the following. The transition probability and the cross section for the proton scattering on an electron gas is given in terms of a four-point CTP Green function in Section \ref{reds}. Section \ref{perts} introduces the lowest order graph for the transition probability and some higher order contributions are identified in Section \ref{highord}. The calculation of the Schwinger-Dyson resummed photon propagator, based on the one-loop self energy expression is outlined in Section \ref{photprops}. Section \ref{scattmds} contains few numerical results and the role of multi-particle final states is clarified in Section \ref{relvirts}. Finally, a brief summary of the results is given in Section \ref{sums}.

\section{Reduction formula}\label{reds}
The initial state of the scattering of a proton beam on an electron gas target is assumed to be
\be
|i\ra=\int\frac{d^3k}{(2\pi)^3}\frac{M}{\Omega_\v{k}}\psi_i(\v{k})a^\dagger_{h_i}(\v{k})|\Psi_g\ra
\ee
where $\psi_i(\v{k})$ is a wave packet of the beam, centered at momentum  $\v{p}$, $a^\dagger_h(\v{k})$ denotes the creation operator of a proton with momentum $\v{k}$ and helicity $h$, $\Omega_\v{k}=\sqrt{M^2+\v{k}^2}$, $M$ being the proton mass, and the state $|\Psi_g\ra$ represents the target, an electron gas. The exclusive transition probability, $P_{exc}=|\la f|S|i\ra|^2$, corresponds to fixed initial and final states and it serves to construct the conditional probability $P^c_{exc}=P_{exc}/|\tilde\psi_i(0)|^2$ where
\be
\tilde\psi_i(\v{x})=\int\frac{d^3k}{(2\pi)^3}\frac{M}{\Omega_\v{k}}e^{i\v{x}\cdot\v{k}}\psi_i(\v{k})
\ee
is the beam wave function in space-time. The usual steps give rise to the expression $P^c_{exc}=|\la f|S|a^\dagger_h(\v{p})|\Psi_g\ra|^2$ for sufficiently narrow wave packet where the momentum dependence of the $S$-matrix is negligible. The inclusive cross section is the average/sum of  $P^c_{exc}/V_4n_g|\v{j}_i|$ over the initial/final states where $V_4$ is the scattering four volume, $n_g$ and $\v{j}_i$ denote the target particle density and beam flux, respectively. The sum here is over all possible final states of the collision process and will be represented as a trace within the Fock space,
\be\label{trpron}
p^c_{inc}=\frac1{V_4}\sum_{h_f}\Tr[a^\dagger_{h_f}(\v{q})a_{h_f}(\v{q})S\rho_iS^\dagger],
\ee
where the $\v{q}$ denotes the momentum of the proton in the final state and the initial density matrix factories as
\be
\rho_i=\hf\sum_{h_i}a^\dagger_{h_i}(\v{p})|\Psi_g\ra\la\Psi_g|a_{h_i}(\v{p}),
\ee
at vanishing temperature, considered in this paper. The covariant expression of the inclusive cross section turns out to be
\be\label{crosss}
\sigma=\frac{mp^c_{inc}}{n_g\sqrt{(p\cdot u)^2-M^2}},
\ee
containing the initial proton momentum $p^\mu=(\Omega_\v{p},\v{p})$ and a four-vector $u$, assuming the form $u^\mu=(1,\v{0})$ in the laboratory frame where the electron gas is at rest.

The calculation of the transition probability is facilitated by the use of the generator functional
\be\label{genfunct}
e^{iW[\hj,\hetab,\heta]}=\Tr\bigl[U(\eta^+,\etab^+,j^+)\rho_iU^\dagger(-\eta^-,-\etab^-,-j^-)\bigr]
\ee
written in units $c=\hbar=1$ where
\be
U(\eta,\etab,j)=T[e^{-i\int_{t_i}^{t_f}dx^0\int d^3x[H-\sum_\tau(\etab_\tau\psi_\tau+\psib_\tau\eta_\tau)-j^\mu A_\mu]}]
\ee
denotes the time evolution operator for the time interval $t_i<t<t_f$ in the presence of the sources $\eta(x),\etab(x),j(x)$ for the charges and the photons, $H(x)$ being the energy density and $\tau=e$ or $p$. We carry out the limit $t_i\to-\infty$, $t_f\to\infty$ when $U\to S$. The path integral representation of the generator functional is
\be\label{wpint}
e^{iW[\hj,\hetab_\tau,\heta_\tau]}=\int D[\hpsi]D[\hpsib]D[\hA]\exp\left\{iS_{CTP}[\hA,\hpsib,\hpsi]+i\sum_\sigma\int dx\left[\sum_\tau(\etab^\sigma_\tau\psi^\sigma_\tau+\psib^\sigma_\tau\eta^\sigma_\tau)+j^{\mu\sigma} A_\mu^\sigma\right]\right\}.
\ee
The presence of $U$ and $U^\dagger$ leads to the reduplication of the fields, $\psi\to\hpsi=(\psi^+,\psi^-)$, etc. in the path integral which satisfy the boundary conditions $\psi^+(t_f,\v{x})=\psi^-(t_f,\v{x})$, $\psib^+(t_f,\v{x})=\psib^-(t_f,\v{x})$ and $A^+(t_f,\v{x})=A^-(t_f,\v{x})$, the incorporation of the trace of Eq. \eq{genfunct}. The CTP action is $S_{CTP}[\hA,\hpsib,\hpsi]=S[A^+,\psib^+,\psi^+]-iS^*[A^-,\psib^-,\psi^-]$, where $S[A,\psib,\psi]$ denotes the action of the electron-proton-photon system.

The conditional transition probability \eq{trpron} can easily be calculated by following the strategy of the reduction formulas,
\bea\label{redform}
P^c_{inc}&=&\frac1{V_2Z_p^2}\int dx^+dy^+dx^-dy^-e^{iq(y^+-y^-)-ip(x^+-x^-)}\nn
&&\times(\ub_fK)^+_{y^+}(\ub_iK)^-_{x^-}(Ku_i)^+_{x^+}(Ku_f)^-_{y^-}\frac{\delta^4W[0,\hetab,\heta]}{\delta\etab_p^-(x^-)\delta\eta_p^-(y^-)\delta\etab_p^+(y^+)
\delta\eta_p^+(x^+)}_{\hetab=\heta=0},
\eea
where $K^\pm=\pm i\partial\br-M$ and its index indicates the argument the space-time derivative is acting upon.

It is worthwhile noting that the generator functional \eq{genfunct} serves as the starting point of Thermal Field Theory \cite{umezawa} (TFT), as well, but the actual calculations in that scheme use
\be\label{tftgenf}
e^{iW_s[\hj,\hetab,\heta]}=\Tr[U(t_f,t_i;\eta^+,\etab^+,j^+)\rho^{1-s}_iU^\dagger(t_f,t_i;-\eta^-,-\etab^-,-j^-)\rho^s_i],
\ee
with $s=1/2$ \cite{niemisa}. Though the generator functional is $s$-independent in equilibrium where $[\rho_i,H]=0$ and $\heta=\hj=0$ the $s$-dependent reappears in a scattering process which is out of equilibrium. The dependence on $s$ is not seen in TFT as long as the observables are inserted either into $U$ or into $U^\dagger$, exclusively. However we have to insert operators into both $U$ and $U^\dagger$ simultaneously in applying the reduction formulae to calculate transition probabilities. Since the field operators which handle the scattering particles appear in between the operators $U^\dagger$ and $\rho_i^s$ in Eq. \eq{tftgenf} do not commute with the initial density matrix one is restricted to the choice $s=0$, which makes TFT equivalent with CTP. The value of $s$ influences the boundary conditions in time only and can be incorporated in the construction of the free propagators within the framework of perturbation expansion. What follows below can be considered as a conventional calculation within TFT by means of a somehow unusual set of free propagators.

\section{Perturbation expansion}\label{perts}
\begin{figure}[ht]
\begin{center}
\begin{picture}(170,40)\includegraphics[scale=.4]{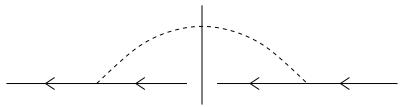}\end{picture}
\caption{Leading order contribution to the transition probability.}\label{ctppq}
\end{center}
\end{figure}

The lowest order, $\ord{e^2}$ graph of $W$ is shown in Fig. \ref{ctppq}. The CTP graphs are constructed according to the standard Feynman rules and the left and right half contain vertices coming from $U$ and $U^\dagger$ in \eq{genfunct}, respectively. These two regions are separated by a vertical line, a cut, in this figure for better visibility. The time runs to the left in both parts, the external particle lines, attached to $U$ or $U^\dagger$ are oriented away from or towards the cut, respectively. The points where the internal lines of the graph which connect $U$ and $U^\dagger$ cross the cut represent the particle content of the state at the final time, contributing to the trace in Eq. \eq{genfunct}. The dashed lines of the graph denotes the free photon propagator in Landau gauge, given by $\hD_0^{\mu\nu}(k)=T^{\mu\nu}(k)\hD_0(k;0)$, where $T^{\mu\nu}(k)=g^{\mu\nu}-k^\mu k^\nu/k^2$ stands for the transverse projector and
\be\label{scctpprop}
\hD_0(k;m)=\begin{pmatrix}\frac{1}{k^2-m^2+i\epsilon}&-2\pi i\delta(k^2-m^2)\Theta(-k^0)\cr
-2\pi i\delta(k^2-m^2)\Theta(k^0)&-\frac{1}{k^2-m^2-i\epsilon}\end{pmatrix}
\ee
is the CTP propagator of a scalar particle of mass $m$ in the basis $(A^+,A^-)$. It can be shown that the CTP two point function for any local bosonic operator is of the form $\hD=C(D^n,D^f,D^i)$ with
\be\label{genprop}
C(C^n,C^f,C^i)=\begin{pmatrix}C^n+iC^i&-C^f+iC^i\cr C^f+iC^i&-C^n+iC^i\end{pmatrix},
\ee
and $D^r=D^n+D^f$, $D^a=D^n-D^f$ are the retarded and advanced Green functions, respectively.
The free proton propagator, $\hG(k)=(k\br+M)\sigma\hD_0(k;M)\sigma$, with
\be
\sigma=\begin{pmatrix}1&0\cr0&-1\end{pmatrix}
\ee
is shown by solid lines. The lines of a CTP graph connecting $U$ and $U^\dagger$ cross a vertical line, placed in the middle of the graph at $t=t_f$, and describe the particle content of final states contributing to the trace  in Eq. \eq{genfunct}.

The inverse free propagators in the reduction formula \eq{redform} remove the proton legs, leaving behind
\be
p^c_{inc}=-ie^2(\ub_f\gamma^\mu u_i)(\ub_i\gamma^\nu u_f)D^{+-}_{0\mu\nu}(r),
\ee
with $r=q-p$. This expression has a simple, suggestive structure, the product of the two interaction vertices and the photon spectral density, given by the Wightman function, $\la0|A_\mu(q-p) A_\nu(p-q)|0\ra$. The averaging (summation) over initial (final) helicities gives
\be\label{lotrprop}
p^c_{inc}=-\frac{ie^2}{2M^2}[p^\mu q^\nu+q^\mu p^\nu+g^{\mu\nu}(M^2-pq)]D^{+-}_{0\mu\nu}(r).
\ee

\section{Higher orders}\label{highord}
\begin{figure}[ht]
\begin{center}
\begin{picture}(150,30)\includegraphics[scale=.4]{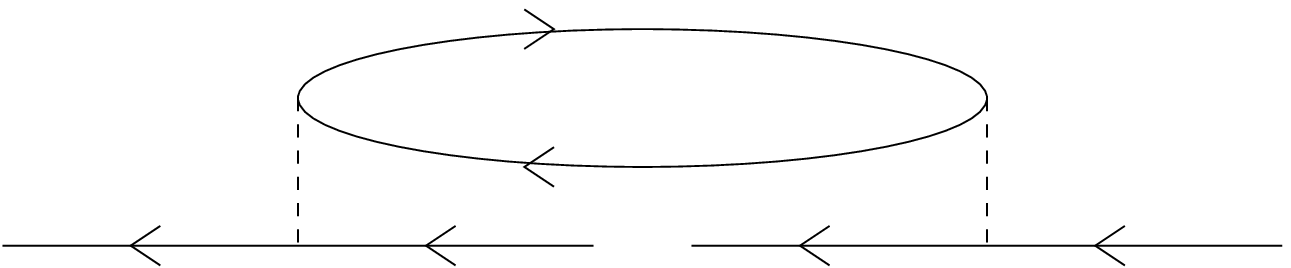}\end{picture}

\begin{center}(a)\end{center}

\begin{picture}(150,30)\includegraphics[scale=.4]{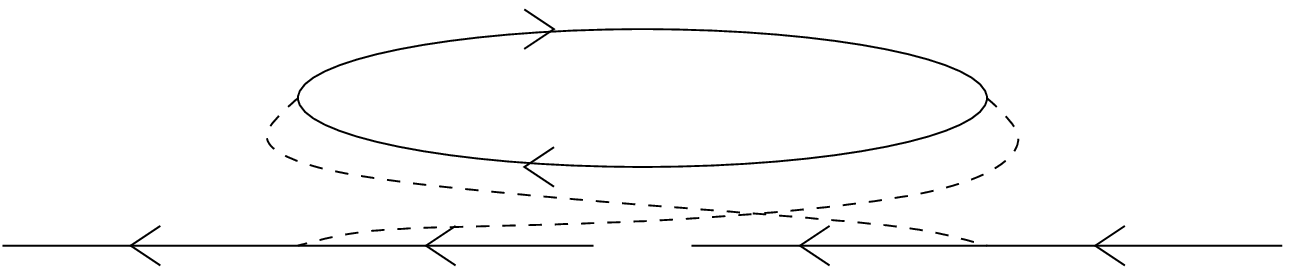}\end{picture}

\begin{center}(b)\end{center}

\caption{$\ord{e^4}$ diagrams of the transition probability of an inelastic proton scattering off an electron gas. (a): CTP diagonal, (b): CTP off-diagonal blocks of the photon propagator is used.}\label{ctppqnho}
\end{center}
\end{figure}

The leading order transition probability \eq{lotrprop} is vanishing because the proton is on-shell before and after the photon emission or absorption, rendering the photon off-shell, $r^2\ne0$, and $D^{+-}_0(r)$ is vanishing in this case. In fact, the cross section starts in order $\ord{e^4}$ hence the non-vanishing contributions to the transition probability must come from higher order graphs. It is easy to find and to resum some of nontrivial contributions by replacing the free photon propagator, $\hD_0$, by
\be\label{schwd}
\hD=\frac1{\hD_0^{-1}-\hat\Pi}
\ee
in the transition probability \eq{lotrprop} where $\hat\Pi$ denotes the photon self energy. The $\ord{e^4}$ order terms where $D_0^{++}\Pi^{+-}D_0^{--}$ or $D_0^{+-}\Pi^{-+}D_0^{+-}$ replaces $D_0^{+-}$ in \eq{lotrprop}, are shown in Fig. \ref{ctppqnho}.

\begin{figure}[ht]
\begin{center}
\begin{picture}(150,30)\includegraphics[scale=.4]{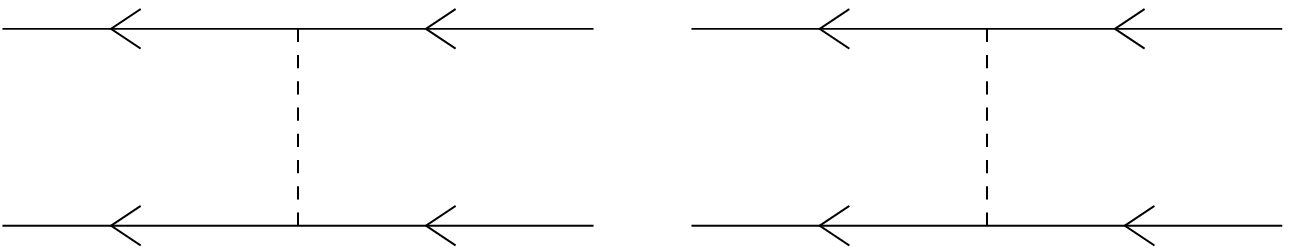}\end{picture}

\begin{center}(a)\end{center}

ls
\begin{picture}(150,40)\includegraphics[scale=.4]{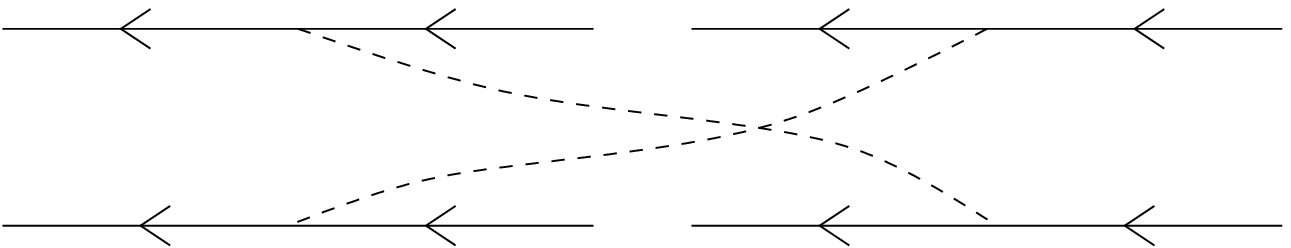}\end{picture}

\begin{center}(b)\end{center}

\caption{Lowest order, $\ord{e^4}$ graphs of the $e+p\to e+p$ elastic transition probability. (a): factorizable, (b): entangled contributions.}\label{ep}
\end{center}
\end{figure}

It is instructive to recall at this point the lowest order graphs of the transition probability of an electron-proton scattering where both the electron and the proton initial and final states are specified \cite{scattenv}, they are depicted in Fig. \ref{ep}, where the upper and lower solid line represent the electron and the proton, respectively. The graphs (b) is vanishing in the same manner as the graph of Fig. \ref{ctppq} but one may use here again the Schwinger-Dyson resummed propagator to recover nontrivial contributions. Graph (a), one photon exchange, is factorizable in contrast to graph (b). Due to the non-factorizability of vertices from $U$ and $U^\dagger$ this latter represents  entanglement among the scattering particles, the system, and the electron gas and photons, the environment.

To recover the transition probability of the process considered in this work we have to sum over the initial and final electron states. This summation can be is achieved in Fig. \ref{ctppqnho} by joining the initial and the final electron legs both at the initial and the final time because the off diagonal CTP block, $G^{-+}$, generated by this step is actually the spectral function. The averaging over the initial states is achieved by the division with the four volume $V_4$ in Eq. \eq{trpron}.

\begin{figure}[ht]
\begin{center}
\begin{picture}(170,80)\includegraphics[scale=.4]{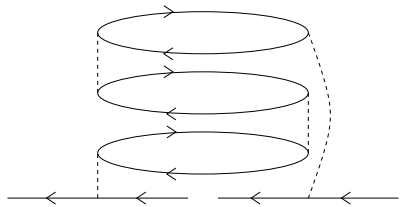}\end{picture}
\caption{Leading order, $\ord{e^6}$ contribution to the graph of Fig. \ref{ctppqnho}(b).}\label{pqeppp}
\end{center}
\end{figure}

The ignorance of the state of one of the colliding particles, the electron, prevents us from identifying the colliding particles-environment entanglement. A diagram represents entanglement if its inner lines connect vertices of $U$ and $U^\dagger$. For instance, Fig. \ref{ctppq} shows proton-electron gas entanglement. But let us now consider the graphs Fig. \ref{ctppqnho}(b). The photon must be in a real electron-hole or electron-positron state at $t=t_f$ because $D_0^{+-}(r)=0$ when $r^2=0$. We shall consider low energy, non-relativistic beam therefore the electron-positron pair contributions will be ignored and the leading order contribution is the graph of Fig. \ref{pqeppp}. If we knew that the electron participating in the collision corresponds to the middle loop of Fig. \ref{pqeppp} then we would consider that loop to be actually made up by external lines and would classify Fig. \ref{pqeppp} as an entangled graph, a higher order contribution to the diagram of Fig. \ref{ep}(b). But it might well be that the colliding electron belongs to the lower loop in Fig. \ref{pqeppp} and in that case we have a higher order correction to the upper, factorizable graph of Fig. \ref{ep}. In simpler words, lacking the proper definition of the colliding particle system we can not define the system-environment entanglement.

This argument shows a characteristic difference between the usual, transition amplitude based formalism and CTP, namely a simple partial resummation of the perturbation series of  a transition probability, obtained in the latter case includes contributions from degenerate multi-particle states which are always present due to the limited resolution power of the experiment. This point of view suggests the separations of graphs representing contributions to transition probabilities into two subsets. An exclusive graph involves the final states, aimed by the experiment only, eg. the graph of Fig. \ref{ctppqnho}(a) since the lines at $t=t_f$ belong to the colliding electron. The rest, the inclusive graphs contain more particles in the final states. The dressing within an exclusive graph takes place within the time axes and is build up in terms of Feynman propagators which contain off-shell, virtual amplitudes. The final states of an inclusive graph are described by the off-diagonal CTP blocks of the propagators, the Wightman function, whose support is in the mass-shell and the corresponding dressing involves real particle states, coupled to both time axes simultaneously.

The closed electron loops of the graphs above are the sum of vacuum and finite temperature/density contributions. We shall consider the electron gas at finite density, parametrized by a non-vanishing Fermi-momentum $k_F\ne0$, at zero temperature. The vacuum, the Dirac-see, and the Fermi-sphere contributions play different dynamical roles in the collision. Imagine that a proton with some momentum is placed into the non-interacting vacuum with vanishing density, $k_F=0$ and go into the rest frame of the proton by performing an appropriate Lorentz transformation. The Dirac-see remains invariant and the proton at rest dresses only, ie. polarizes the electron Dirac-see when the interactions are turned on. The virtual vacuum-polarizations stay around the proton and generate no collision, eg. the vacuum contribution to the graph of Fig. \ref{ctppqnho}(a) is vanishing. In other words, the possibility to make the electron Dirac-see co-moving with the proton excludes scattering and such dressing is generated by exclusive graphs. But at finite density when the Lorentz boost to the proton rest frame ends up in an excited electron state the proton enters into a non-trivial collision process and changes its momentum as soon as the interactions are turned on and inclusive graph contributions emerge.

\section{Photon propagator}\label{photprops}
The CTP structure of the self energy can be found by noting that the inverse of the block matrix \eq{genprop} is of the form $C^{-1}(C^n,C^f,C^i)=\sigma C(C'^n,C'^f,C'^i)\sigma$. For instance, the inverse of the free scalar propagator \eq{scctpprop},
\be\label{invfctppr}
\hD^{-1}_0(k;m)=(k^2-m^2)\begin{pmatrix}1&0\cr0&-1\end{pmatrix}
+i\epsilon\begin{pmatrix}1&-2\Theta(-k^0)\cr-2\Theta(k^0)&1\end{pmatrix},
\ee
what yields $D_0^{-1n}=k^2-m^2$, $D_0^{-1i}=\epsilon$ and $D_0^{-1f}=i\mr{sign}(k^0)\epsilon$ can be found by using the regulated Dirac-delta $\delta_\epsilon(z)=\epsilon/(z^2+\epsilon^2)\pi$ in the inversion. Such a structure allows us to write the photon self energy as $\hat\Pi=\sigma C(\Pi^n,\Pi^f,\Pi^i)\sigma$ and to define the retarded, advanced and imaginary parts of the self energy, $\Pi^r=\Pi^n+\Pi^f$, $\Pi^a=\Pi^n-\Pi^f$ and $\Pi^i$. The Schwinger-Dyson resummed photon propagator yields $D^{\stackrel{r}{a}}=[D_0^{-1}-\Pi^{\stackrel{r}{a}}]^{-1}$ and $D^i=D^r\Pi^iD^a$ where $D_0^{-1\mu\nu}(r)=-T^{\mu\nu}(r)r^2$ \cite{maxwell}. These equations can be summarized as $\hD=D^r\sigma\hD^{-1\dagger}\sigma D^a$, in particular
\be\label{dpmsum}
\hD^{+-}=D^r\hat\Pi^{+-}D^a.
\ee
The key factor of the transition probability \eq{lotrprop}, $\hD^{+-}$, the photon spectral weight is given as the spectral function of particle-hole excitations, $\Pi^{+-}$ weighted by the retarded and advanced photon propagator.

The Lorentz structure of symmetric, transverse tensors depending on the four-vectors $r$ and $u$ contains two independent scalars and can be parametrized in an $O(3)$ covariant manner as
\be
P^{\mu\nu}(c_\ell,c_t)=\frac{c_\ell}{1-\nu^2}\begin{pmatrix}1&\v{n}\nu\cr\v{n}\nu&\nu^2\v{L}\end{pmatrix}-c_t\begin{pmatrix}0&0\cr0&\v{T}\end{pmatrix},
\ee
where $r^\mu=(\omega,\v{r})$, $\nu=\omega/|\v{r}|$, $\v{n}=\v{r}/|\v{r}|$, $\v{L}=\v{n}\otimes\v{n}$ and $\v{T}=1-\v{L}$. It is easy to find the inverse within the transverse subspace, $(c_\ell,c_t)\to(1/c_\ell,1/c_t)$, since $P^{\mu\rho}(c_\ell,c_t)P^{~\nu}_\rho(1/c_\ell,1/c_t)=T^{\mu\nu}$. It is actually simpler to calculate the combinations $\hat\Pi_g=\hat\Pi^{\mu\nu}g_{\mu\nu}$, $\hat\Pi_u=u\hat\Pi u$ with the CTP structure
\bea\label{olselen}
\Pi^n_x(r)&=&\pi^+_x(\omega,\v{r})+\pi^+_x(-\omega,\v{r}),\nn
\Pi^f_x(r)&=&i[\pi^-_x(\omega,\v{r})-\pi^-_x(-\omega,\v{r})],\nn
\Pi^i_x(r)&=&\pi^i_x(\omega,\v{r})+\pi^i_x(-\omega,\v{r}),
\eea
where the index $x$ stands either for $g$ or $u$. The rule of converting the parametrization $\hat\Pi_g$, $\hat\Pi_u$ into $\hat\Pi_\ell$, $\hat\Pi_t$ is $\Pi_\ell=(1-\nu^2)\Pi_u$ and $\Pi_t=[\Pi_u-(1-\nu^2)\Pi_g]/2$.

The one-loop expression for a non-degenerate electron gas, $k_F\ll m$, $m$ being the electron mass at vanishing temperature is \cite{maxwell}
\bea\label{pisf}
\pi^+_x&=&2Cf_xL,\nn
\pi^-_x&=&-\pi Cf_xM,\nn
\pi^i_x&=&-\pi Cf_xN,
\eea
with $C=\alpha k_F^2m/\pi|\v{r}|$, $f_g=1+r^2/2m^2$, $f_u=1+r^2/4m^2+\omega/m$ and
\bea\label{reimxf}
L&=&z+\hf(1-z^2)\ln\left|\frac{z+1}{z-1}\right|,\nn
M&=&\Theta(1-|z|)(1-z^2),\nn
N&=&\begin{cases}1-z^2&|\v{r}|>2k_F,~-1<z<1\cr
1-z^2&|\v{r}|<2k_F,~1-\frac{|\v{r}|}{k_F}<z<1\cr
\frac{(\omega+2m)\omega}{k_F^2}&|\v{r}|<2k_F,~-\frac{|\v{r}|}{2k_F}<z<1-\frac{|\v{r}|}{k_F}\end{cases},
\eea
with $z(r)=(r^2+2m\omega)/2|\v{r}|k_F$. Note that $L$ not to be confused with the spatial longitudinal projector $\v{L}$. The form $D^{-1}_0(r)=-r^2P(1,1)$ of the free photon propagator and \eq{olselen} lead to $D_\ell^{\stackrel{r}{a}}=1/(\v{r}^2-\Pi_u^{\stackrel{r}{a}})(1-\nu^2)$ and $D_t^{\stackrel{r}{a}}=2/[(2\v{r}^2+\Pi_u^{\stackrel{r}{a}})(1-\nu^2)-\Pi_g^{\stackrel{r}{a}}]$. These expressions agree with well known results, namely $\Pi^n_u$ is the Lindhard function and $\Pi^f_u+i\Pi^i_u$ is the spectral function of particle-hole pairs of non-relativistic electron gas. The preceding equations, together with the generic matrix element
\be
pP(D_\ell,D_t)q=D_\ell\frac{(q^0-\v{n}\v{q}\nu)(p^0-\v{p}\v{n}\nu)}{1-\nu^2}-D_t[\v{p}\v{q}-(\v{p}\v{n})(\v{n}\v{q})],
\ee
where $p^\mu=(p^0,\v{p})$, $q^\mu=(q^0,\v{q})$ give the cross section
\be
\sigma=-\frac{ie^2}{|\v{p}|Mn_g}\biggl[D^{+-}_\ell\frac{(q^0-\v{n}\v{q}\nu)(p^0-\v{p}\v{n}\nu)}{1-\nu^2}-D^{+-}_t[\v{p}\v{q}-(\v{p}\v{n})(\v{n}\v{q})]+\hf(M^2-p^0q^0+\v{p}\v{q})(D^{+-}_\ell+2D^{+-}_t)\biggr],
\ee
where the  electron density can be approximated by its noninteracting form, $n_g=k_F^3/3\pi^2$.

\begin{figure}[ht]
\begin{center}
\begin{picture}(170,140)\includegraphics[scale=.5]{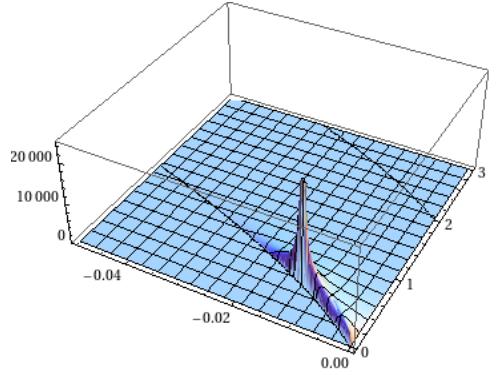}\end{picture}
\caption{The spectral function $\Im D^{+-}_\ell$, plotted on the plane $(\omega/k_F,|\v{r}|/k_F)$, for $-0.05k_F<\omega<0<|\v{r}|<3k_F$.}\label{dpm}
\end{center}
\end{figure}

\section{Scattering at metallic density}\label{scattmds}
Let us now consider few plots of a scattering which  takes place in an electron gas with $k_F=0.02m$, chosen to be around metallic density. First we look into the electric spectral function, $\Im D^{+-}_\ell$, shown in Fig. \ref{dpm}, the transverse spectral function, $\Im D^{+-}_t$  being negligible in the same region. The electric spectral function is the product of two factors, $\Im D^{+-}_\ell(r)=-|D^r_\ell(r)|^2\Im\Pi^{+-}_\ell(r)$, according to Eq. \eq{dpmsum}. In the one-loop approximation $-i\Pi^{+-}$ is the spectral weight of a single particle-hole excitation and both $\Pi_\ell^f(r)$ and $\Pi_\ell^i(r)$ and are non-vanishing when $|z(r)|<1$ or $|z(-r)|<1$, according to Eqs. \eq{pisf}-\eq{reimxf}. These inequalities are satisfied within the strip
\be\label{phphsp}
(|\v{r}|-k_F)^2-k_F^2<\omega^2+2m|\omega|<(|\v{r}|+k_F)^2-k_F^2
\ee
of the plane $(|\v{r}|,\omega)$. We have $|\omega|\ll m$ in the region displayed therefore the support of the spectral function is approximately between two parabolas on the plane $(|\v{r}|,\omega)$, given by the inequalities \eq{phphsp} with $\omega^2$ ignored, and indicated in Fig \ref{dpm} by two lines drawn on the plane $(\omega/k_F,|\v{r}|/k_F)$ at zero altitude. The higher-loop contributions to the self energy become small but non-vanishing beyond the strip \eq{phphsp}.

\begin{figure}[ht]
\begin{center}
\begin{picture}(170,160)\includegraphics[scale=.45]{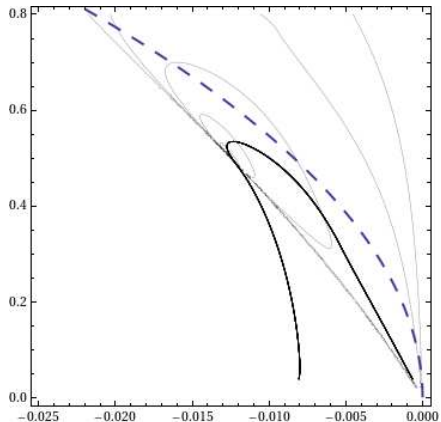}\end{picture}
\end{center}
\caption{Solid line: Electron-hole quasi-particle dispersion relation on the plane $(\omega/k_F,|\v{r}|/k_F)$, for $-0.025k_F<\omega<0$ and $0.03k_F<|\v{r}|<0.8k_F$. Thin lines: The contour plot of spectral function $\Im D^{+-}_\ell$. Dashed line: The curve obtained by varying $p$ with $q=0.99988p$ and $\theta=0.4\cdot10^{-4}\pi$.}\label{dpmc}
\end{figure}

The other factor in $\Im D^{+-}_\ell$, the retarded propagator, assumes large values in the vicinity of the dispersion relation of the electron-hole quasi-particles, $\Re D^{r-1}(\omega,\v{r})=0$, and where the life time of the excitations is long, $\Im D^{r-1}(\omega,\v{r})$ is small. The quasi-particles, the plasmon and the zero-sound modes correspond to the vertical and the leftward bending part of the thick line on Fig. \ref{dpmc}, respectively .The thin solid lines make up the contour-plot of the spectral function $\Im D^{+-}_\ell$, c.f. Fig. \ref{dpm} and $\Im D^{+-}_\ell=0$ under the lowest, approximately diagonal contour line. Locate the singular point $(\omega_s,r_s)$ on Fig. \ref{dpmc} where the quasi-particle line leaves the support of $\Pi^{+-}_\ell$. The scattering is non-perturbative in the vicinity of the two-dimensional sphere $S$ in the photon four-momentum, made up by the points $(\omega_s,\v{r})$ where $|\v{r}|=r_s$.

We have $|\Pi^f|\ll\Pi^i$ within the kinematical region in question where due to Eqs. \eq{pisf}-\eq{reimxf}
\be
-i\Pi^f\sim\left(1+\frac{\omega}{m}\right)\left(1-\frac{m^2\omega^2}{k_F^2\v{r}^2}\right)-(\omega\to-\omega),
\ee
and $-i\Pi^{+-}=i\Pi^f+\Pi^i\sim i\Pi^f$ forms a ridge approximately along the line
\be
\omega=\frac{|\v{r}|k_F}{\sqrt3m}
\ee
which becomes sharp as $\omega\to0$. The zero-sound line follows this line at small wave vectors but the increase of $|\v{r}|$ bends it towards higher frequencies compared to this linear dispersion relation due to the flattening of the rim.

Though both plasmons and zero-sound modes are quasi-particles they are not equally important. The importance of an electron-hole pair in the scattering as an intermediate state is measured by $\Im D^{+-}_\ell$. This function is determined by three factors, namely the distance from the ``mass-shell'' $\Re D^{r-1}$, the life-time $1/\Im D^{r-1}$ and the density of states $-i\Pi^{+-}$. Plasmons, though being long living excitations of the electron gas, are suppressed by the spectral function according to Fig. \ref{dpmc} and they appear in the collision through the higher-loop photon self-energy only. The scattering is dominated by the vicinity of the sphere $S$ where the plasmon and the zero-sound lines merge.

\begin{figure}[ht]
\begin{picture}(450,130)
\includegraphics[scale=.35]{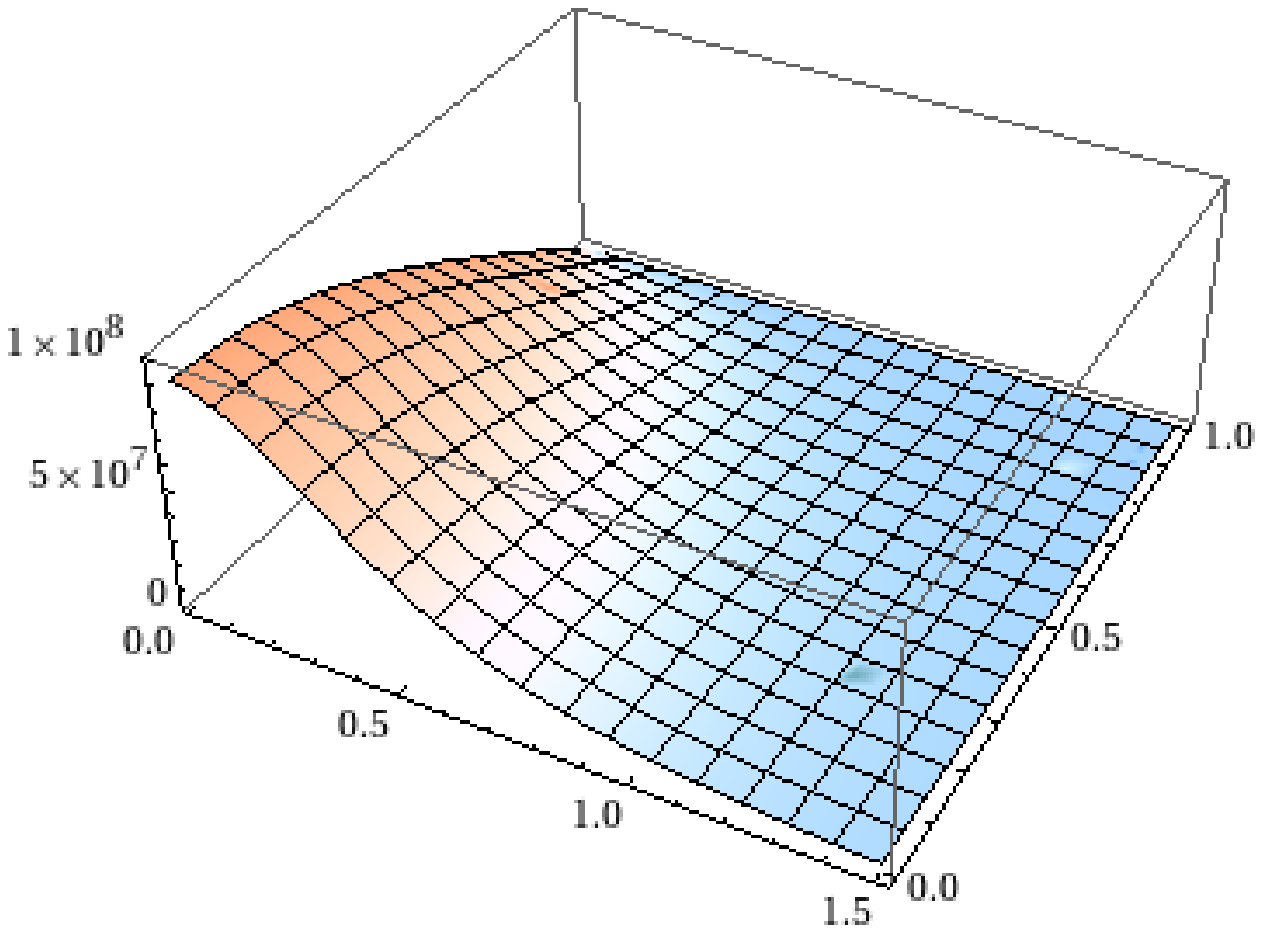}
\includegraphics[scale=.35]{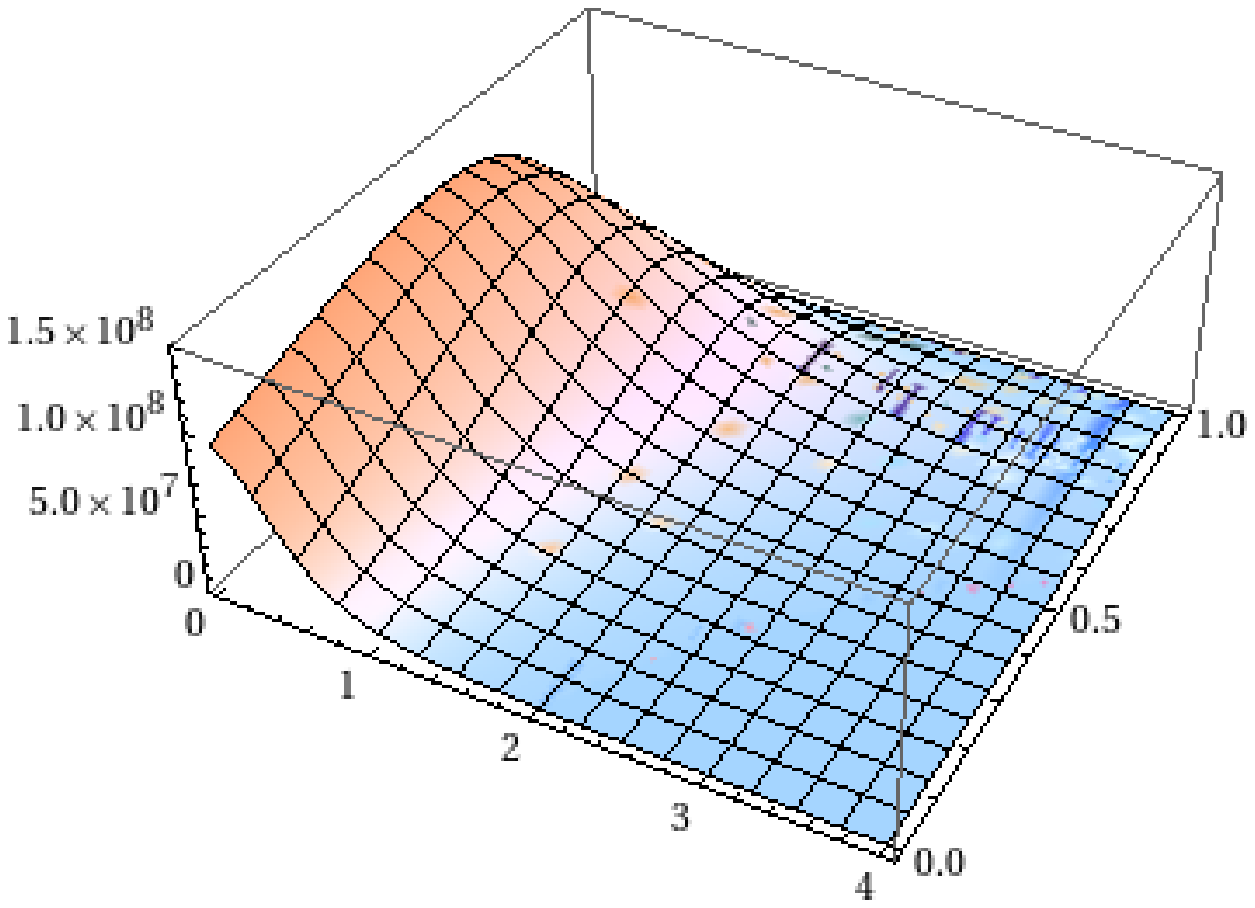}
\includegraphics[scale=.35]{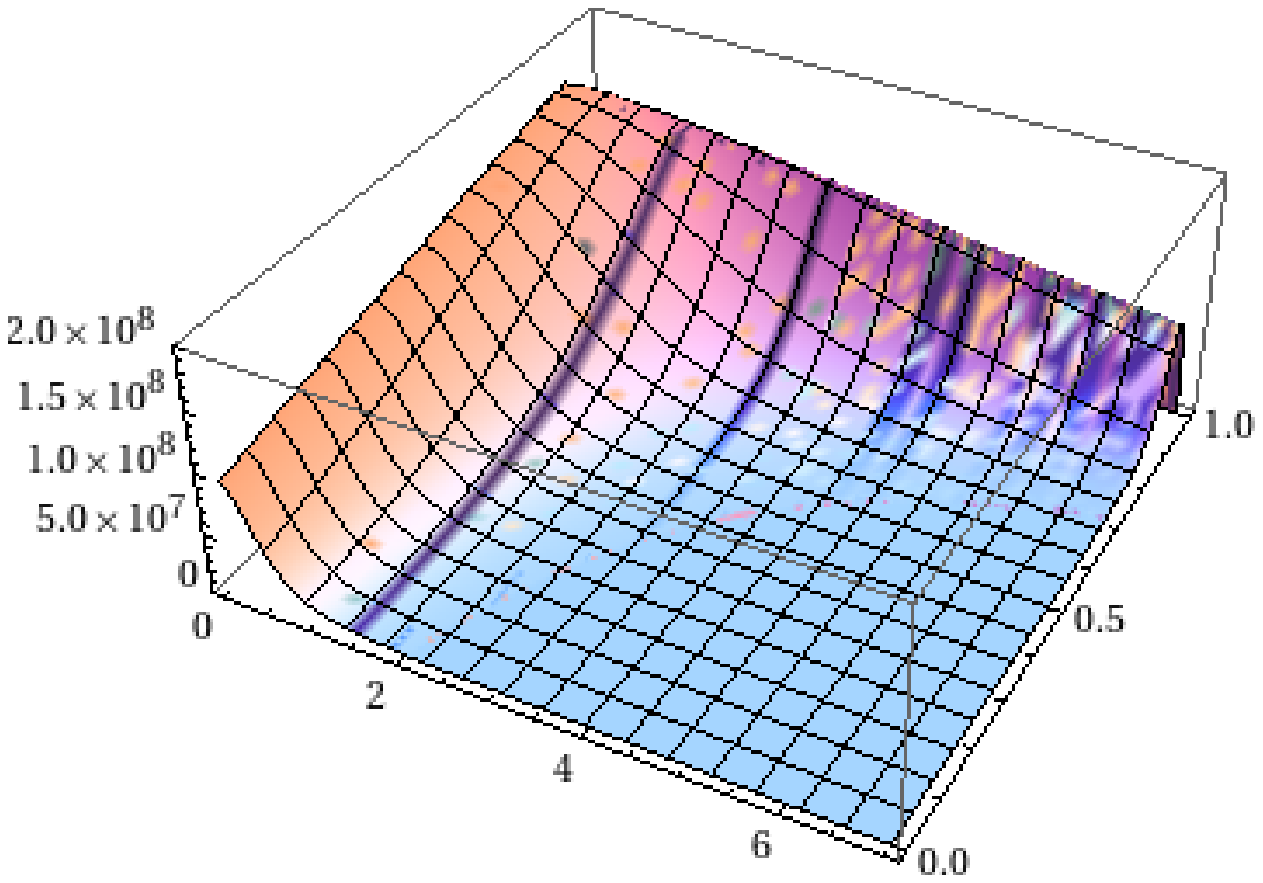}
\end{picture}
\begin{center}(a)\hskip5.5cm (b)\hskip5.5cm (c)\end{center}
\caption{Cross section plotted on the plane $(p/k_F,q/p)$ for (a): $\theta=0.5\pi$, (b): $\theta=0.1\pi$ and (c): $\theta=0.01\pi$.}\label{sigp}
\end{figure}

The cross section \eq{crosss} where the colliding electron is not recorded corresponds to inelastic proton scattering which will characterized by three independent parameters, the initial and final proton momentum, $p$ and $q<p$, respectively and the scattering angle $\theta$, all considered in the rest frame of the target. The cross section is shown in Fig. \ref{sigp} for three different values of the scattering angle. It displays weak variation for $0.5\pi<\theta<\pi$ but its peak starts to drift toward larger $q$ values as $\theta$ is decreased below $0.5\pi$. The peak approaches $q=p$ and a sharply localized rim develops at $q$ values slightly below the elastic limit, $q=p$ in the forward scattering limit, $\theta\to0$. The cross section is always vanishing for $q=p$.

It is worthwhile mentioning that the integral of the cross section in $q$, calculated numerically approaches the elastic proton-electron cross section in the vacuum as $k_F\to0$.

\section{Real vs. virtual excitations}\label{relvirts}
We turn now to the weight of inclusive graphs in the cross section. For this end we split the self energy into the sum of CTP-diagonal and non-diagonal terms, $\hat\Pi=\hat\Pi_d+\hat\Pi_{nd}$, and introduce $\hD_d=[\hD_0^{-1}-\hat\Pi_d]^{-1}$, including the CTP-diagonal one-loop self energy corrections, the electron-hole dressing completely within either the operator $U$ or $U^\dagger$ of \eq{genfunct}. The direct inversion of the $2\times2$ CTP matrix given by Eq. \eq{invfctppr} and $\hat\Pi_d=Diag[\Pi^n+i\Pi^i,-\Pi^n+i\Pi^i]$ leads to
\be\label{dhdsdrs}
\hD_d(r)=\begin{pmatrix}\frac1{-r^2+i\epsilon-\Pi^n-i\Pi^i}&-\frac{2i\epsilon\Theta(-\omega)}{(r^2+\Pi^n)^2+(\Pi^i-\epsilon)^2}\cr-\frac{2i\epsilon\Theta(\omega)}{(r^2+\Pi^n)^2+(\Pi^i-\epsilon)^2}&\frac1{r^2+i\epsilon+\Pi^n-i\Pi^i}\end{pmatrix}.
\ee
Note that the off-diagonal blocks are vanishing if their denumerator is finite, when either $\Pi^i(r)$ or $r^2+\Pi^n(r)$ is non-vanishing. Apart of the quasi-particle line of
Fig. \ref{dpmc} $r^2+\Pi^n(r)\ne0$ but the quasi-particle life-time is finite, giving $\Pi^i(r)\ne0$ on this line. Therefore the propagator \eq{dhdsdrs} is CTP-diagonal, $\hD_d=Diag[D_F,-D_F^*]$ and the Feynman propagator $D_F$, appearing at this point is dressed by the one-loop self energy, $D_F=[D_0^{-1}-\Pi^n-i\Pi^i]^{-1}$, $\Pi^n$ and $\Pi^i$ being given by \eq{olselen}-\eq{reimxf}.

The CTP propagator \eq{schwd} can be brought into the form
\be
\hD=\frac1{\hD_d^{-1}-\hat\Pi_{nd}},
\ee
showing clearly the separation of the graphs into exclusive and inclusive classes. In fact, let us consider the CTP block
\be\label{dpmseries}
\hD^{+-}=\sum_{n=0}^\infty[\hD_d(\hat\Pi_{nd}\hD_d)^{2n+1}]^{+-},
\ee
appearing in the transition probability \eq{lotrprop}. The $\ord{\hat\Pi_{nd}}$ part is the exclusive contribution,
\be
D^{+-}_{exc}=[\hD_d\hat\Pi_{nd}\hD_d]^{+-}=D_F(i\Pi^i-\Pi^f)D_F^*,
\ee
because the cross section $\sigma_{exc}$ obtained by replacing $D^{+-}_0$ by $D^{+-}_{exc}$ in the transition probability \eq{lotrprop} corresponds a single electron in the final state. The transition probability \eq{lotrprop} is non-negative because $-i\Im D^{+-(r)}_y\ge0$, $y=\ell,t$ is the photon spectral function. This inequality applies to any value of electric charge $e$ therefore it holds for each order of the perturbation expansion. The exclusive contribution, $-i\Im D^{+-(r)}_{exc~y}$, the spectral weight of a single photon decaying into a real electron-hole pair is therefore positive definite because Fig. \ref{ctppqnho}(a) is the only $\ord{e^2}$ piece in the photon density of state.

The inclusive contributions to the transition probability come from the higher order terms in the series \eq{dpmseries} since each graph contributing to $D^{+-}_{inc}=\hD^{+-}-\hD_{exc}^{+-}$ contains several electron-hole pairs at the final time. Multi-particle graphs can be visualized by expanding the photon line in the self energy in Fig. \ref{ctppq} and retaining more than one electron-hole self energy insertion crossing the vertical line, i.e. connecting $U$ and $U^\dagger$, as in Fig. \ref{pqeppp}. An $\ord{\hat\Pi_{nd}^{2n+1}}$ graph can be split into $n$ blocks of two electron and two holes, more precisely two electron-hole pairs with opposite energy-momentum. Charge conjugation invariance makes possible that the electron and the hole lines belong to real excitations. Such self-energy insertions do not cover complete orders of the loop or perturbation expansion and therefore their contribution to the density of state is not definite.

The relation $(\hat\Pi_{nd}\hD_F)^2=D_F(\Pi^{f2}+\Pi^{i2})D^*_F$ allows us to carry out the sum \eq{dpmseries} with the result
\be
\hD^{+-}=\frac{D_F(-\Pi^f+i\Pi^i)D_F^*}{1-D_F(\Pi^{f2}+\Pi^{i2})D^*_F},
\ee
yielding the ratio
\be\label{ctpnctpr}
\frac{D_{excl~y}^{+-}}{\hD^{+-}_y}=1-|D_{F~y}|^2(\Pi_y^{f2}+\Pi_y^{i2}).
\ee
Since $\Pi^f(r)$ is imaginary and $\Pi^i(r)$ is real $\Pi_y^{f2}+\Pi_y^{i2}=-|\Pi_y^f|^2+\Pi_y^{i2}$ and the inclusive contributions to the cross section is positive or negative for $|\Pi_y^f|<|\Pi_y^i|$ or $|\Pi_y^f|>|\Pi_y^i|$, respectively. We have $|\Pi_y^f|>|\Pi_y^i|$ in our case as mentioned above and a destructive interference is formed between the exclusive and the inclusive diagrams.

\begin{figure}[ht]
\begin{center}
\begin{picture}(320,130)\includegraphics[scale=.45]{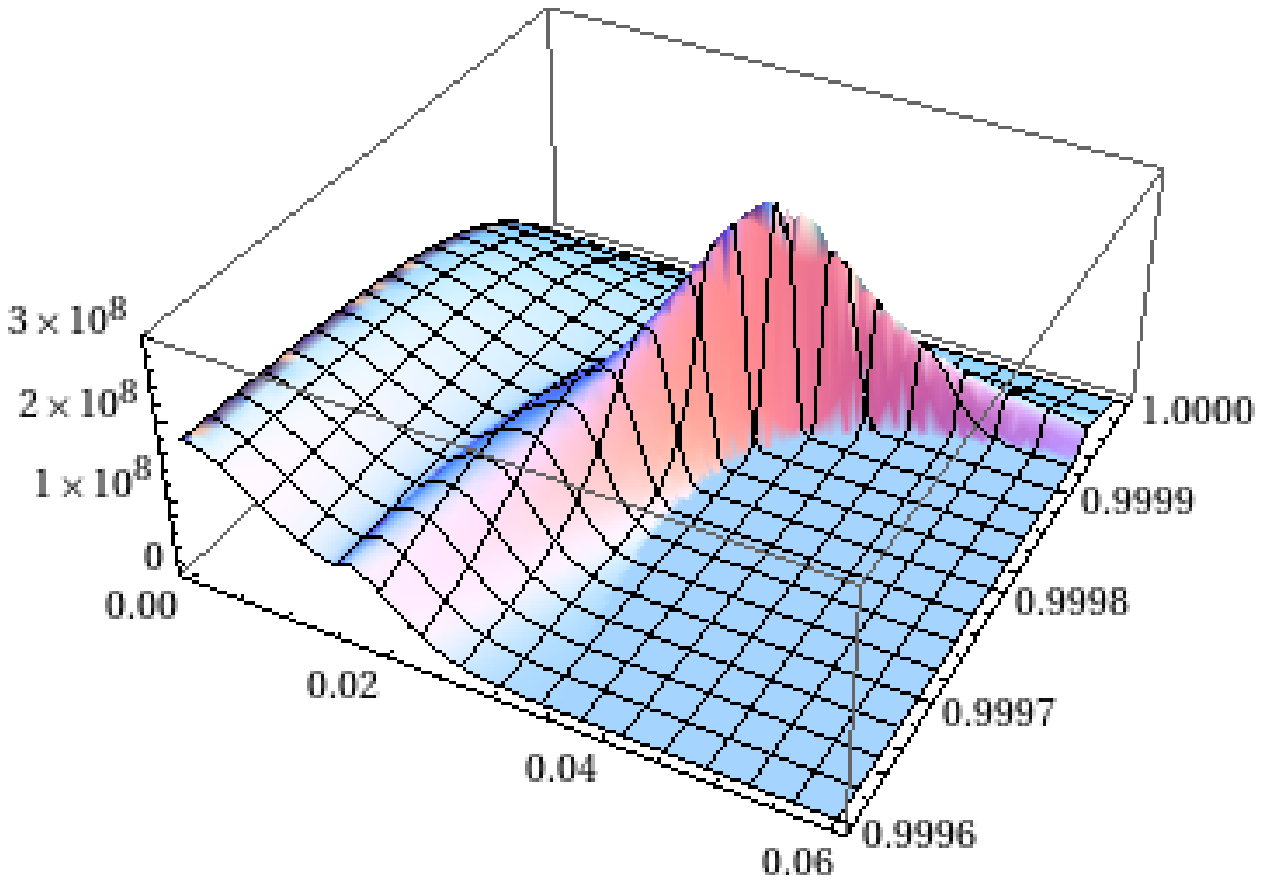}\hskip1cm
\includegraphics[scale=.45]{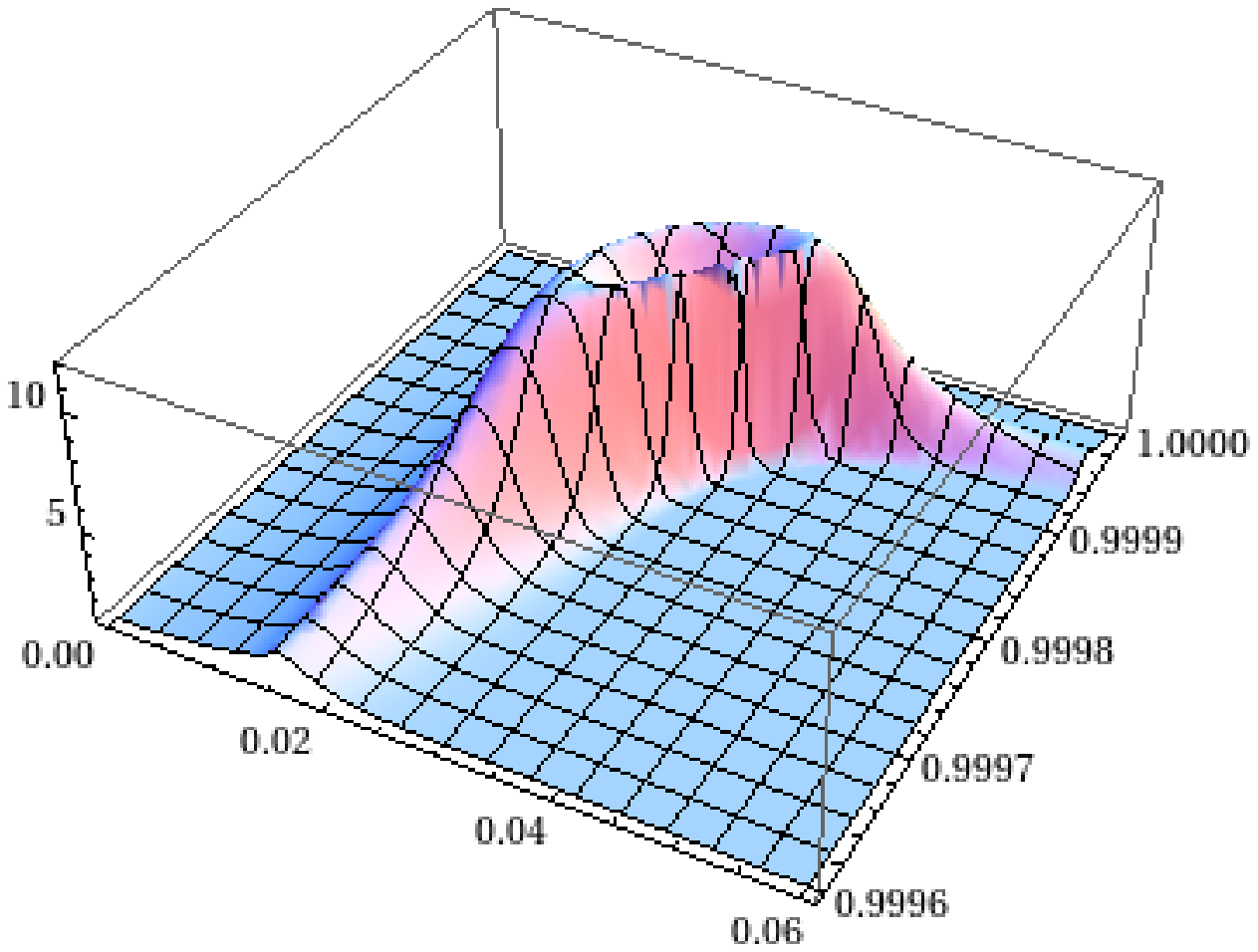}\end{picture}

\begin{center}(a)\hskip5.5cm (b)\end{center}
\caption{(a): The cross section $\sigma$, (b): the ratio $\sigma_{exc}/\sigma$ plotted on the plane $(p/M,q/p)$ for $\theta=0.4\cdot10^{-4}\pi$.}\label{sigrat}
\end{center}
\end{figure}

In order to reach the non-perturbative kinematical regime, the two-dimensional sphere $S$, we need large proton momentum and small scattering angle. The collision is nearly elastic, $q\approx p$, close to forward scattering according to the last figure of \ref{sigp} and the cross section displays a rim with a peculiar peak as the scattering angle is decreased as shown in Fig. \ref{sigrat}(a). The ratio $\sigma_{exc}/\sigma$ displays first a peak at the same location which turns into a crater for weaker scattering angle, as shown in Fig. \ref{sigrat}(b).

\begin{figure}[ht]
\begin{center}
\begin{picture}(170,130)\includegraphics[scale=.4]{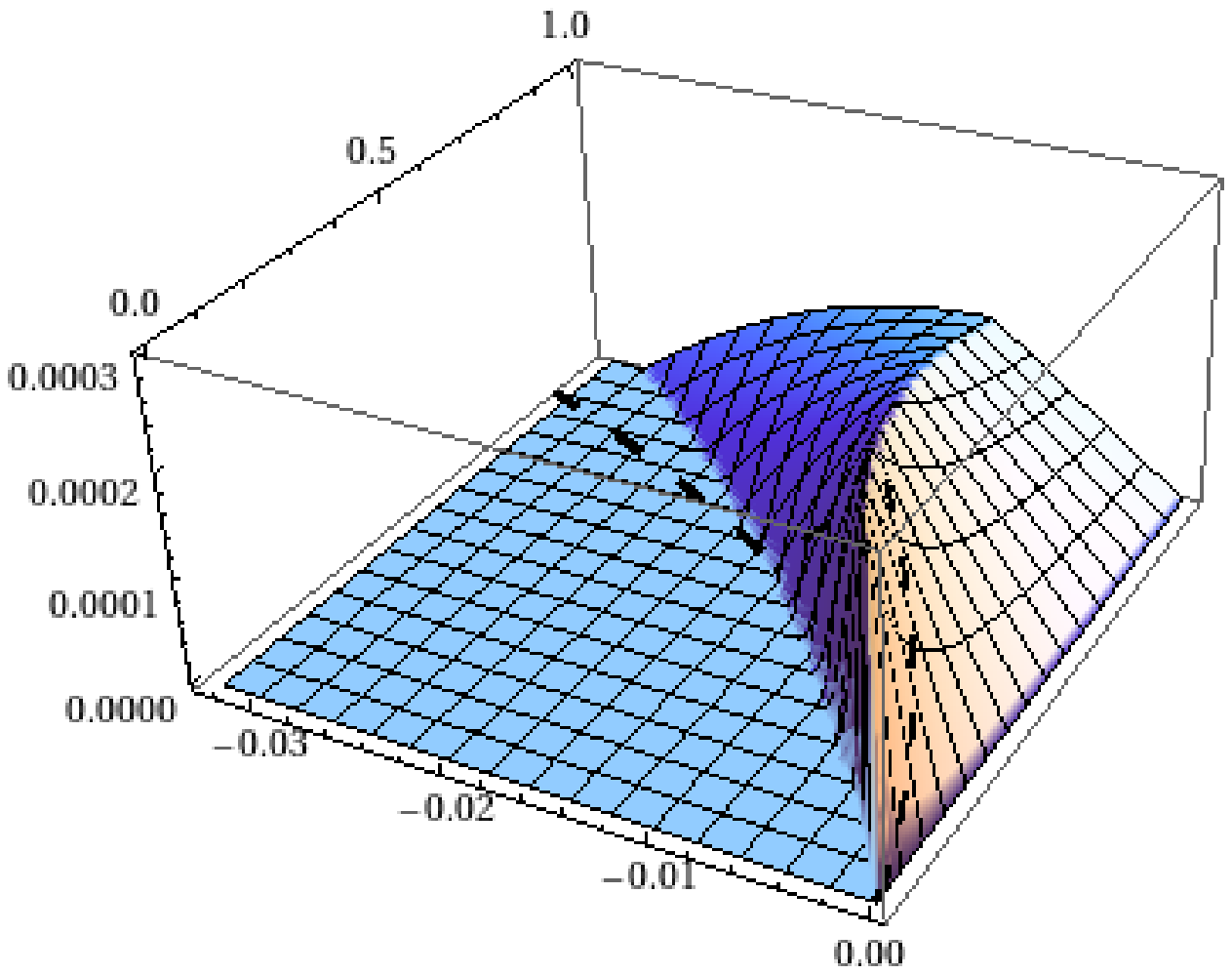}\end{picture}\begin{picture}(170,130)\includegraphics[scale=.4]{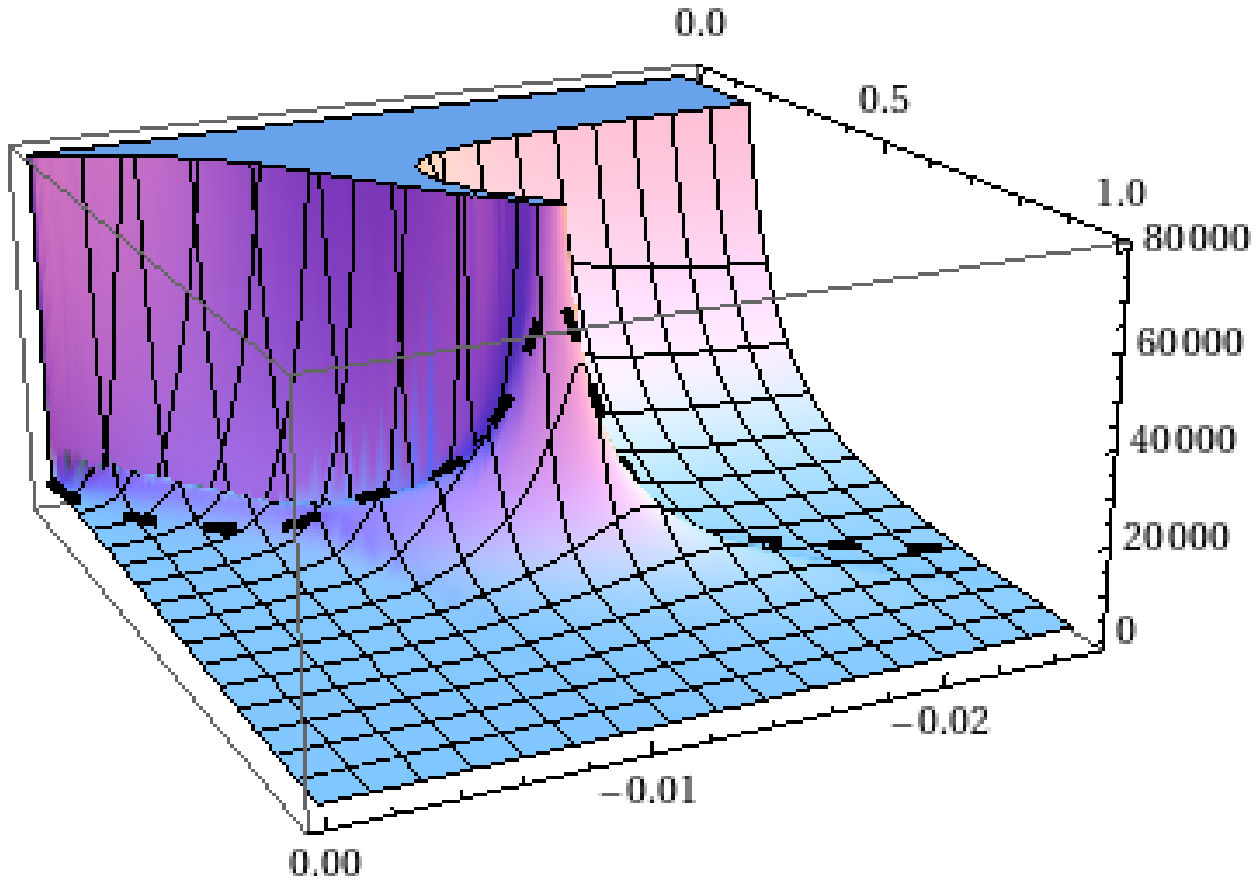}\end{picture}

\begin{center}(a)\hskip5.5cm (b)\end{center}
\caption{The factor (a): $\sqrt{\Pi_y^{f2}+\Pi_y^{i2}}$ and (b): $|D_F|$, plotted on the plane $(\omega/k_F,|\v{r}|/k_F)$ for $-0.032k_F<\omega<0$, $0<|\v{r}|<k_F$. The dashed curve represents the line $q=0.99988p$ and $\theta=0.4\cdot10^{-4}\pi$.}\label{sigratl}
\end{center}
\end{figure}

To understand better the strong cancellation between the exclusive and the inclusive graphs, the large values of $\sigma_{exc}/\sigma$, both factors of the ratio \eq{ctpnctpr} are plotted in Fig. \ref{sigratl}. The line of $\theta=0.4\cdot10^{-4}\pi$ and $q=0.99988p$, varying $p$ is displayed as a dashed thick line in Figs. \ref{dpmc} and \ref{sigratl}. It climbs through the sharp ridge of the geometric mean of the density of states of electron-hole pairs with four-momentum $r$ or $-r$ in the final states according to Fig. \ref{sigratl}(a). Fig. \ref{sigratl}(b) shows that $|D_F|$ assumes large values as the line approaches the two-dimensional sphere $S$ in Fig. \ref{dpmc}. The crater of a depth approximately $20\%$ of the maximum of $\sigma_{exc}/\sigma$ is due to the higher orders in $D_{inc}^{+-}$ because $D^{+-}_{exc}$ displays no particular structure around the peak of $\sigma_{exc}$.

The strong cancellation among the exclusive and inclusive graphs is reminiscent of the cancellation IR divergences among real and virtual soft photons \cite{bloch,yennie}. The soft photon contributions are potentially dangerous and require special care because the photon propagator representing them is large. It is the large value of $|D_F|^2$ what drives the cancellation in our case, too. The exclusive graph of Fig. \ref{ctppqnho}(a) assumes large values when the factor $|D_F|^2$, arising from the two photon lines is large, when the energy-momentum of the photon is close to the two-dimensional sphere $S$. The electron-hole pairs appearing in $D_F$ are virtual because they belong either to $U$ or $U^\dagger$ rather than the final state. On the contrary, the electron-hole pairs of $\Pi^{\pm\mp}$ appear in the final state and are real. The cancellation between the exclusive and inclusive contributions, driven by electron-hole pairs is strong for $1\ll|D_{F~y}|^2|\Pi_y^{f2}+\Pi_y^{i2}|$. It takes place when real electron-hole pairs occupy high density of state and virtual photons spend a large fraction of time as a virtual electron-hole pair.

\section{Summary}\label{sums}
It was shown that a part of the final state interactions of scattering processes, namely excitations made by the collision on a many-body system in equilibrium can be accounted of in a natural and simple manner within the CTP formalism. The perturbation expansion gives rise a graphical representation of expectation values without relying on auxiliary, non-observable transition amplitudes. As a result, CTP graphs allow us to treat diagonal and interference terms of the transition probability on equal footing.

Another advantage of the graphical structure of the CTP perturbation series is the flexible handling of the final state. One subset of graphs, called here exclusive contributions, represents the dressing of the colliding particles only. This contribution is the absolute magnitude square of the transition amplitude and sums up the impact of the environment on the collision ignoring the back-reaction, the change in the environment state. The rest, called inclusive contribution stands for pieces of the transition probability which belong to final states with additional, real excitations of the environment.

The transition probability was based on the lowest order graph in this work but a partial resummation of the perturbation series was carried out by placing the one-loop self energy in the denominator of the photon propagator. Each Feynman photon propagator appearing in exclusive and in inclusive graphs comes with the same four-momentum in this approximation and becomes large, rendering the collision non-perturbative when the four-momentum, exchanged between the colliding particles is close to the region where the plasmon and zero-sound lines merge.

A strong cancellation was found between the exclusive and inclusive graphs for non-perturbative collisions. It is similar to the cancellation of IR divergences in QED except that it is now a cancellation between large but finite contributions of virtual and real electron-hole pairs. The problem of the IR  divergences was not addressed in this work because they cancel at finite density and vanishing temperature according to the general picture \cite{kinoshita,lee}, in the case considered here and should be absent at finite temperature, as well \cite{weldon}. But it remains to be seen if there is a similarly strong cancellation among the finite contributions, as found in the present work, when vertex function and all self energy insertions are retained.

One expects that the present scheme of scattering processes can address realistic problems by including the environment and should be pursued in plasma and astrophysical applications. As it stands, this scheme describes the cross section which can be defined within a plasma. When a beam scatters off a target then one has to include the container of the target in the description. This naturally changes the qualitative details of the cross section but the interesting effects, namely the enhancement of the cross section by the collective modes within the plasma and the strong cancellation between real and virtual excitations remain present.

\end{document}